\documentclass{article}
\usepackage{spconf,amsmath,graphicx}
\usepackage[ruled,vlined]{algorithm2e}
\usepackage{hyperref}
\usepackage{subfig}
\usepackage{color,soul}
\usepackage{multirow}
\usepackage{cite}
\usepackage{siunitx}
\usepackage{stfloats} %to get the figure* not on last page
    
\graphicspath{ {./images/} }

\makeatletter
\let\origsection\section
\renewcommand\section{\@ifstar{\starsection}{\nostarsection}}
\let\origssection\subsection
\renewcommand\subsection{\@ifstar{\starssection}{\nostarssection}}
\newcommand\nostarsection[1]
{\sectionprelude\origsection{#1}\sectionpostlude}
\newcommand\starsection[1]
{\sectionprelude\origsection*{#1}\sectionpostlude}
\newcommand\nostarssection[1]
{\sectionprelude\origssection{#1}\sectionpostlude}
\newcommand\starssection[1]
{\sectionprelude\origssection*{#1}\sectionpostlude}
\newcommand\sectionprelude{%
  \vspace{-0.5em}
}
\newcommand\sectionpostlude{%
  \vspace{-0.2em}
}
\makeatother

\newcommand{\dynbox}[2][c]{\begin{tabular}{@{}#1@{}}#2\end{tabular}}

\DeclareMathOperator*{\argmax}{arg\,max}

\title{A Deep Learning based Pipeline for Efficient Oral Cancer Screening on Whole Slide Images}

\name{\dynbox{Jiahao Lu$^{\star}$,
Nata{\v s}a Sladoje$^{\star}$, 
Christina Runow Stark$^{\mathsection}$, \\ 
Eva Darai Ramqvist$^{\ddagger}$, 
Jan-Micha{\'e}l Hirsch$^{\mathparagraph}$, 
Joakim Lindblad$^{\star}$}\thanks{This work is supported by: Swedish Research Council proj. 2015-05878 and 2017-04385,
VINNOVA grant 2017-02447, FTV Stockholms L{\"a}n AB}}
\address{$^{\star}$Centre for Image Analysis, Dept. of Information Technology, Uppsala University, Sweden\\
$^{\mathsection}$Dept. of Orofacial Medicine at S{\"o}dersjukhuset, Folktandv{\aa}rden Stockholms L{\"a}n AB, Sweden\\
$^{\ddagger}$Karolinska Universitetsjukhuset, Dept. of Clinical Pathology and Cytology, Stockholm, Sweden\\
$^{\mathparagraph}$Department of Surgical sciences, Uppsala University, Sweden
}
\begin{document}
\maketitle              % typeset the header of the contribution

\begin{abstract}
Oral cancer incidence is rapidly increasing worldwide. The most important determinant factor in cancer survival is early diagnosis. To facilitate large scale screening, we propose a fully automated pipeline for oral cancer detection on whole slide cytology images. The pipeline consists of fully convolutional regression-based nucleus detection, followed by per-cell focus selection, and CNN based classification. %On the same samples, it is shown to outperform classification of human-selected nucleus patches. 
Our novel focus selection step provides fast per-cell focus decisions at human-level accuracy. 
We demonstrate that the pipeline provides 
%fast and 
efficient cancer classification of whole slide cytology images, improving over previous results both in terms of accuracy and feasibility. 
The complete source code is made available as open source\footnote{\url{https://github.com/MIDA-group/OralScreen}}.
\end{abstract}
\section{Introduction}
\label{sec:intro}

%Cancer of the oral cavity is one of the most common malignancies in the world \cite{stewartWorldCancerReport2019}. Similar to cervical cancer, visual inspection of brush collected samples has shown to be a practical and effective approach to reduce mortality \cite{sankaranarayananLongTermEffect2013}.
%
%However, considering costs involved, no country has yet introduced a national screening program for oral cancer \cite{speightScreeningOralCancer2017}. We aim to reduce the prevalence of oral cancer by introducing screening of high risk patients in General Dental Practice by dentists and dental hygienists. Computer assisted cytological examination is essential for feasibility of this project.

Cancers in the oral cavity or the oropharynx are among the most common malignancies in the world \cite{stewartWorldCancerReport2019,warnakulasuriya2009global}. 
Similar as for cervical cancer, visual inspection of brush collected samples has shown to be a practical and effective approach for early diagnosis and reduced mortality \cite{sankaranarayananLongTermEffect2013}. 
%To reduce the prevalence of oral cancer, 
We, therefore, work towards introducing screening of high risk patients in General Dental Practice by dentists and dental hygienists. Computer assisted cytological examination is essential for feasibility of this project, due to large data and high involved costs\cite{speightScreeningOralCancer2017}.
%
%However, considering costs involved, no country has yet introduced a national screening program for oral cancer \cite{speightScreeningOralCancer2017}. 

Whole slide imaging (WSI) 
refers to scanning of conventional microscopy glass slides to produce digital slides. 
WSI
%This imaging modality 
is gaining popularity among pathologists worldwide, due to its potential to improve diagnostic accuracy, increase workflow efficiency, and improve integration of images into information systems\cite{farahani2015whole}. Due to the very large amount of data produced by WSI, typically generating images of around 100,000$\times$100,000 pixels with up to 100,000 cells, manipulation and analysis are challenging and require special techniques. %Compression is often employed to enable storage of the acquired data. 
In spite of these challenges, the advantage to reproduce the traditional light microscopy experience in digital format
%, offered by virtual slides on a digital screen, 
makes WSI a very appealing choice.  

Deep learning (DL) %, the most powerful approach in modern image processing, 
has shown to perform very well in cancer classification.
%different classification tasks using Whole Slide Images (WSIs) \cite{campanellaClinicalgradeComputationalPathology2019, cruz-roaAccurateReproducibleInvasive2017, korbarDeepLearningClassification2017, teramotoAutomatedClassificationBenign2019, kirangvAutomaticClassificationWhole2019}. 
An important advantage, compared to (classic) model-based approaches, is absence of need for nucleus segmentation, a difficult task typically required for otherwise subsequent feature extraction. At the same time, the large amount of data provided by WSI makes DL a natural and favorable choice.
In this paper we present a complete fully automated DL based segmentation-free pipeline for oral cancer screening on WSI.

\section{Background and Related work}
\label{sec:background}

A number of studies suggest to use DL for classification of histology WSI samples,
%^Deep learning has shown to perform well in different classification tasks using Whole Slide Images (WSIs)
\cite{campanellaClinicalgradeComputationalPathology2019, cruz-roaAccurateReproducibleInvasive2017, korbarDeepLearningClassification2017, teramotoAutomatedClassificationBenign2019}. 
A common approach is to split tissue WSIs into smaller patches and perform analysis on the patch level.
%\cite{kirangvAutomaticClassificationWhole2019}. 
Cytological samples are, however, rather different from tissue. For tissue analysis the larger scale arrangement of cells is important and region segmentation and processing is natural. For cytology, though, the extra-cellular morphology is lost and cells are essentially individual (especially for liquid based samples); the natural unit of examination is the cell. 

Cytology generally has slightly higher resolution requirements than histology; texture is very important and accurate focus is therefore essential. On the other hand, auto-focus of slide scanners works much better for tissue samples being more or less flat surfaces. In cytology, cells are partly overlapping and at different z-levels. Tools for tissue analysis rarely allow z-stacks (focus level stacks) or provide tools for handling such. In this work we present a carefully designed complete chain of processing steps for handling cytology WSIs acquired at multiple focus levels, including cell detection, per-cell focus selection, and CNN based classification.

Malignancy-associated changes (MACs) are subtle morphological changes that occur in histologically normal cells due to their proximity to a tumor.
%and may include changes in the size, shape, and chromatin structure of the nucleus\cite{palcic1994nuclear}. 
MACs have been shown to be reproducibly measured via image cytometry for numerous cancer types\cite{us2005malignancy}, making them potentially useful as diagnostic biomarkers. 
%Cytotechnologists classification of oral cancer is based on the subtle variation in nucleus texture
Using a random forest classifier \cite{jabaleeIdentificationMalignancyAssociatedChanges2018} reliably detected MACs in histologically normal (normal-appearing) oropharyngeal epithelial cells located in tissue samples adjacent to a tumor and suggests to use the approach as a noninvasive means of detecting early-stage oropharyngeal tumors. 
%Probably no space for further discussion:
%Their feature selection procedure confirms earlier observations that texture features are most significant for reliable detection.
Reliance on MAC enables using patient-level diagnosis for training of a cell-level classifier, where \textit{all} cells of a patient are assigned the same label (either cancer or healthy)\cite{forslidDeepConvolutionalNeural2017}. This hugely reduces the burden of otherwise very difficult and laborious manual annotation on a cell level.

%classification of oral cancer is based on the subtle variation in nucleus texture and a classifier on a cell level is expected to achieve better or similar performance \cite{houPatchBasedConvolutionalNeural2016}. 
%Our here proposed pipeline is based on generation of cell patches through automatic detection of individual nuclei from oral WSIs. This includes nucleus detection and focus selection.

{\bf Cell detection:} 
State-of-the-art object detection methods, such as the R-CNN family\cite{girshickRichFeatureHierarchies2014, girshickFastRCNN2015, renFasterRCNNRealTime2016} and YOLO\cite{redmonYouOnlyLook2016, redmonYOLO9000BetterFaster2017, redmonYOLOv3IncrementalImprovement2018}, have shown satisfactory performance for natural images. However, being designed for computer vision, where perspective changes the size of objects, we find them not ideal for cell detection in microscopy images. Although appealing to learn end-to-end the classification directly from the input images, s.t. the network jointly learns region of interest (RoI) selection \textit{and} classification, for cytology WSIs this is rather impractical. The classification task is very difficult and requires tens of thousands of cells to reach top performance, while a per-cell RoI detection is much easier to train (much fewer annotated cell locations are needed), requires less detail and can be performed at lower resolution (thus faster). To jointly train localization and classification would require the (manual) localization of the full tens of thousands of cells. Our proposal, relying on patient-level annotations for the difficult classification task, reaches good performance using only around 1000 manually marked cell locations. Methods for detecting objects with various size and the bounding boxes also cost unnecessary computation, since all cell nuclei are of similar size and bounding box is not of interest in diagnosis. Further, these methods tend to not handle very large numbers of small and clustered objects very well\cite{zouObjectDetection202019}. 

Many DL-based methods specifically designed for the task of nucleus detection are similar to the framework summarized in \cite{hofenerDeepLearningNuclei2018}: first generate a probability map by sliding a binary patch classifier over the whole image, then find  nuclei positions as local maxima. However, considering that WSIs are as large as 10 giga-pixels, this approach is prohibitively slow.
%, where high performance usually requires small step-size.
U-Net models %can also be applied, 
avoid the sliding window and reduce computation time. Detection is 
performed as segmentation where each nucleus is marked as a binary disk \cite{falkUNetDeepLearning2019}. However, when images are noisy and with densely packed nuclei, the binary output mask
%being like a segmentation focusing on edges 
is not ideal for further processing. 
We find the regression approach \cite{xieMicroscopyCellCounting2018, kainzYouShouldUse2015, xieClassificationStructuredRegression2015}, where the network is trained to reproduce fuzzy nuclei markers, to be more appropriate.
%This approach provides an end-to-end inference for arbitrary image size by fully convolutional networks.

{\bf Focus selection:} 
%Good focus is essential for cancer classification. 
In cytological analysis, the focus level has to be selected for each nucleus individually, since different cells are at different depth. Standard tools (e.g., the microscope auto-focus) fail since they only provide a large field-of-view optimum, and often focus on clumps or other artifacts. 
%The subsequent focus selection step can be formulated as an image quality assessment problem. 
%Many methods have been suggested to provide no-reference blur assessment.
%for natural images. 
Building on the approaches of Just Noticeable Blur (JNB) \cite{ferzliNoReferenceObjectiveImage2009} and Cumulative Probability of Blur Detection (CPBD) \cite{narvekarNoReferenceImageBlur2011}, the Edge Model based Blur Metric (EMBM) \cite{guanNoreferenceBlurAssessment2015} provides a no-reference blur metric by using a parametric edge model to detect and describe edges with both contrast and width information. It claims to achieve comparable results to the former while being faster.

% Just Noticeable Blur (JNB) \cite{ferzliNoReferenceObjectiveImage2009} provides a blur metric by integrating a perceptual blur concept with a probability summation model. Cumulative Probability of Blur Detection (CPBD) \cite{narvekarNoReferenceImageBlur2011} estimates and accumulates the probability of each edge being detected. Edge Model based Blur Metric (EMBM) \cite{guanNoreferenceBlurAssessment2015} uses a parametric edge model to detect and describe edges with both contrast and width information.
% %of each edge pixel. 
% This metric claims to achieve comparable results to JNB and CPBD while being faster. 
% %However, in microscopy images of oral cells, the out-of-focus problem causes not only blur but also artefacts with sharp edges in the background, which could be confusing for the above methods. Here we made some improvements to resolve the above problem.

% \noindent
{\bf Classification:} %The final step of the proposed pipeline is classification. 
Deep learning has successfully been used for different types cell classification\cite{guptaDeepLearningImage2019} and for cervical cancer screening in particular\cite{zhangDeepPapDeepConvolutional2017}. Convolutional Neural Networks (CNNs) have shown ability to differentiate healthy and malignant cell samples\cite{forslidDeepConvolutionalNeural2017}. 
%To reduce labor and avoid human bias, Ground Truth (GT) labels are in \cite{forslidDeepConvolutionalNeural2017} defined  only at the patient level, not at the cell level.
Whereas the approach in \cite{forslidDeepConvolutionalNeural2017} 
relies on manually selected free lying cells, our study proposes to use automatic cell detection.
This allows improved performance by scaling up 
%By automated cell detection this study presents a method which 
the available data to \textit{all} free lying cells in each sample.
%, thanks to efficient and fast detection step.

%Using a random forest classifier \cite{jabaleeIdentificationMalignancyAssociatedChanges2018} reliably detected Malignancy-associated changes (MAC) in histologically normal (normal-appearing) oropharyngeal epithelial cells located in tissue samples adjacent to a tumor and could be used as a noninvasive means of detecting early-stage oropharyngeal tumors. 
%Probably no space for further discussion:
%Their feature selection procedure confirms earlier observations that texture features are most significant for reliable detection.

%\cite{kirangvAutomaticClassificationWhole2019}

\section{Material and Methods}
\label{sec:method}

%Our proposed pipeline consists of regression based nucleus detection followed by per-cell focus selection and finally CNN based classification. 

\subsection{Data}
\label{sec:ExperimentalSetup_data}

%Three datasets are used to develop and evaluate this pipeline. 
Three sets of images of oral brush samples are used in this study. 
{\bf Dataset 1} is a relatively small Pap smear dataset imaged with a standard microscope.
{\bf Dataset 2} consist of WSIs of the same glass slides as Dataset 1.
{\bf Dataset 3} consist of WSIs of liquid-based (LBC) prepared slides.
All samples are collected at Dept.\ of Orofacial Medicine, Folktandv{\aa}rden Stockholms l{\"a}n AB. 
From each patient, samples were collected with a brush scraped at areas of interest in the oral cavity. Each scrape was either smeared onto a glass (Datasets 1 and 2) or placed in a liquid vial (Dataset 3). All samples were stained with standard Papanicolau stain. Dataset 3 was prepared with Hologic T5 ThinPrep Equipment and standard non-gynecologic protocol.
\textbf{Dataset 1} was imaged with an Olympus BX51 bright-field microscope with a $20\times$, 0.75 NA lens giving a pixel size of \SI{0.32}{\um}.
From 10 Pap smears (10 patients), free lying cells (same as in ``Oral Dataset 1'' in \cite{forslidDeepConvolutionalNeural2017}) are  manually selected and $80\!\times\!80\!\times\!1$ grayscale patches are extracted, 
%The number of cells per glass ranges from 226 to 1965. 
each with one centered in-focus cell nucleus.
\textbf{Dataset 2}: The same 10 slides as in Dataset 1 were imaged using a NanoZoomer S60 Digital slide scanner, $40\times$, 0.75 NA objective, at
%multiple focus levels. 
%It contains 10 NDPI files, each containing whole slide microscopy images of oral cells of 7 magnification levels (0.625x, 1.25x, 2.5x, 5x, 10x, 20x, and 40x) 
 11 z-offsets ($\pm$\SI{2}{\um}, step-size \SI{0.4}{\um}) providing RGB WSIs of size $103936\!\times\!107520\!\times\!3$, \SI{0.23}{\um}/pixel. \textbf{Dataset 3} was obtained in the same way as Dataset 2, but from 12 LBC slides from 12 other patients.

%In each dataset, there are only weak labels on glass level. 
Slide level annotation and reliance on MAC appears as a useful way to avoid need for large scale very difficult manual cell level annotations. Both \cite{jabaleeIdentificationMalignancyAssociatedChanges2018} and \cite{forslidDeepConvolutionalNeural2017} demonstrate promising results for MAC detection in histology and cytology. In our work we therefore aim to classify cells based on the patient diagnosis, i.e., all cells from a patient with diagnosed oral cancer are labeled as cancer.

\subsection{Nucleus Detection}
\label{sec:method_ND}

The nucleus detection step aims to efficiently detect each individual cell nucleus in WSIs. The detection is inspired by the Fully Convolutional Regression Networks (FCRNs) approach proposed in \cite{xieMicroscopyCellCounting2018} for cell counting.
\begin{figure}[tbp]
    \centering
    \subfloat[Original image, $I$]{\includegraphics[width=0.45\textwidth]{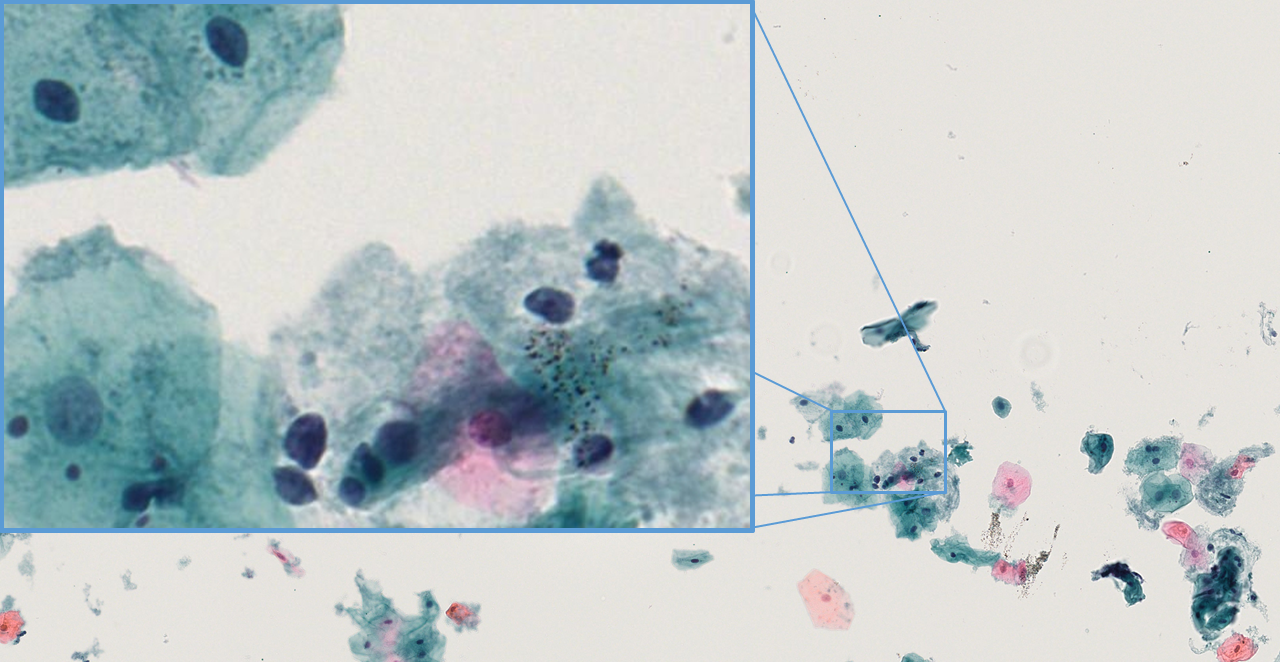}\label{fig:original}}\hfil
    \subfloat[Fuzzy ground truth, $D$]{\includegraphics[width=0.45\textwidth]{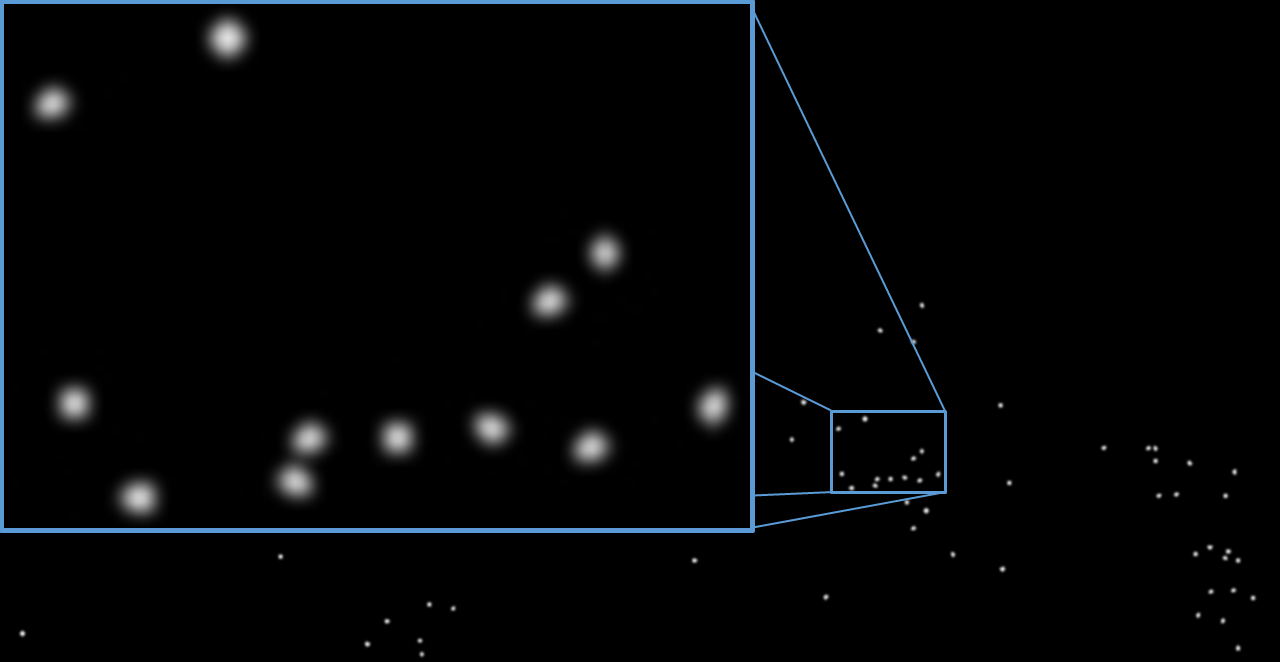}\label{fig:groundtruth}}
    
    \subfloat[Predicted density map $D'\;$\; \text{(in pseudo color)}]{     \includegraphics[width=0.45\textwidth]{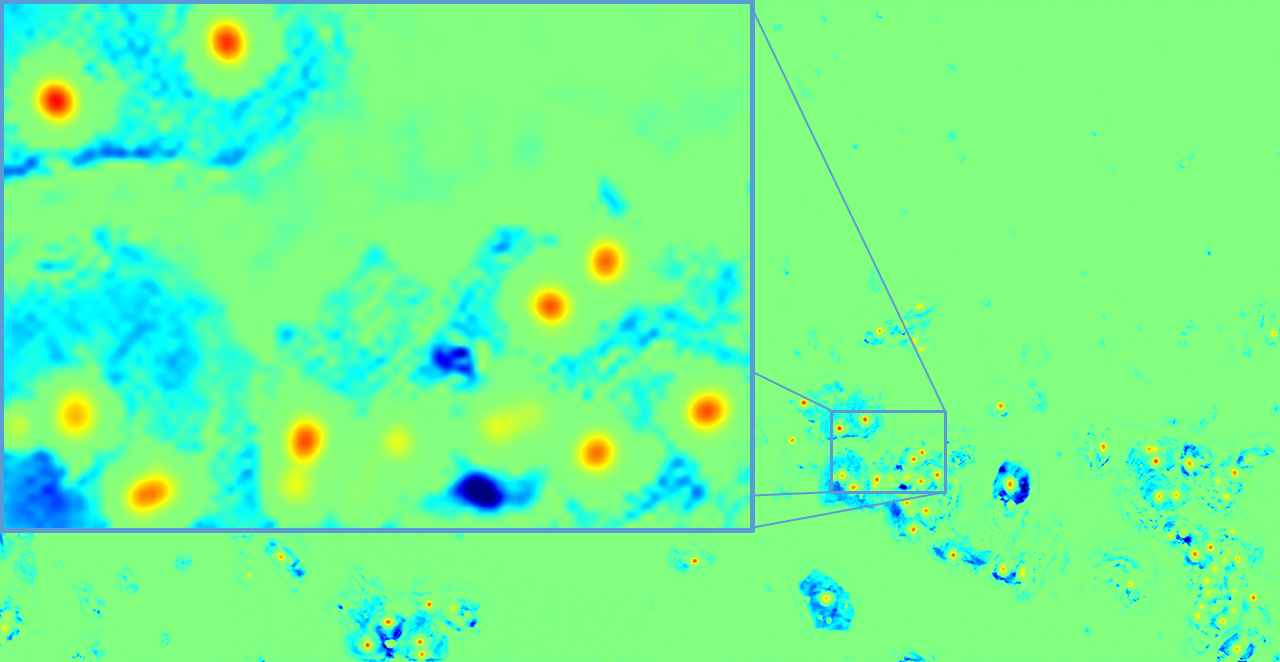}\label{fig:pmap}}\hfil
    \subfloat[\!Detected (blue $\times$) and true (green~+) nuclei locations
    %after thresholding
    % (blue -- detection, green -- ground truth)
    ]{\includegraphics[width=0.45\textwidth]{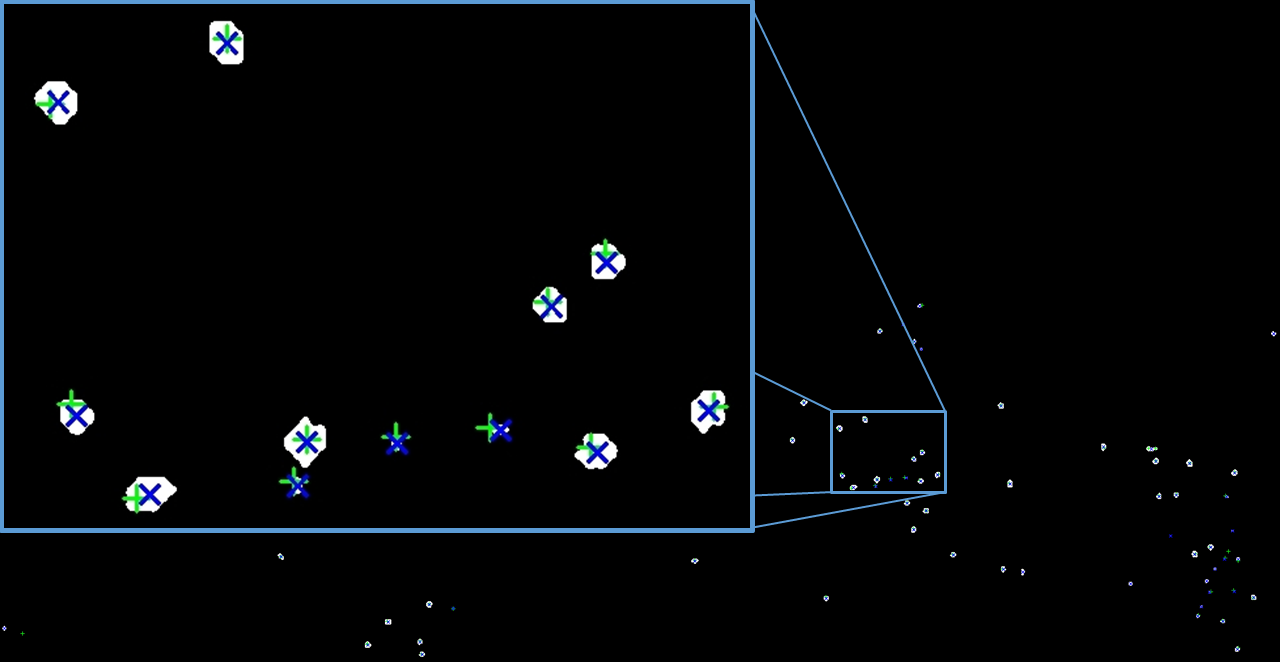}\label{fig:prediction}}
    
    \caption{A sample image at different stages of nucleus detection\!}
    \label{fig:sample}
\end{figure}
The main steps of the method are described below, and illustrated on an example image from Dataset 3, Fig.~\ref{fig:sample}. 

\noindent \textbf{Training:}
Input is a set of RGB images $I_i, i = 1\ldots K$, and corresponding binary annotation masks $B_i$, where each individual nucleus is indicated by one centrally located pixel.

%\begin{figure}[tbp]
%    \centering
%    \includegraphics[width=0.45\textwidth]{images/U-Net.png}
%    \caption{U-Net architecture}
%    \label{fig:unet}
%\end{figure}
%
%\begin{itemize}
  %  \item 
%\noindent \textbf{Pre-processing}: 
%Morphological dilation by a binary disk of radius $r$ is applied on each ground truth mask 
Each ground truth mask is dilated by a disk of radius $r$
\cite{falkUNetDeepLearning2019}, followed by 
convolution % of the output 
with a 2D Gaussian filter of width $\sigma$. 
By this, a fuzzy ground truth is generated.
   %:
    %$$ f(x, y)=e ^ {-\left({\frac %{x^2+y^2}{2\sigma^{2}}}\right)}, $$
    %where $\sigma$ is the spread of the blob.
%\item 
%
%\noindent \textbf{Training}: 
A fully convolutional network is trained to learn a mapping between the original image $I$ (Fig.~\ref{fig:original}) and the corresponding ``fuzzy ground truth'', $D$ (Fig.~\ref{fig:groundtruth}). The network follows the architecture of U-Net \cite{ronnebergerUNetConvolutionalNetworks2015} but with the final softmax replaced by a linear activation function.
    %, as shown in Figure \ref{fig:unet}.
%    
%\item 

\noindent \textbf{Inference:} A 
corresponding 
density map $D'$ (Fig.~\ref{fig:pmap}) is generated (predicted) for any given test image $I$.
%
% \item 
%
%\noindent \textbf{Post-processing}: 
The density map 
$D'$ %resulting from the inference 
is thresholded at a level $T$
%back to a binary mask 
%where each pixel value larger than the threshold $T$ is set to one otherwise zero, 
and centroids of the resulting blobs indicate detected nuclei locations (Fig.~\ref{fig:prediction}).
%\end{itemize}

\subsection{Focus Selection}
\label{sec:method_FC}

%The focus selection step aims to select a good focus for each detected nucleus. 
% The focus level has to be selected for each nucleus individually, since different cells are at different depth. Standard tools (e.g., the microscope auto-focus) fail since they often focus on clumps or other artifacts.
%
Slide scanners do not provide sufficiently good focus for cytological samples and a focus selection step is needed.
Our proposed method utilizes $N$ equidistant z-levels acquired of the same specimen. Traversing the z-levels, the change between consecutive images shows the largest variance at the point where the specimen moves in/out of focus.
This novel focus selection approach provides a clear improvement over the Edge Model based Blur Metric (EMBM) proposed in\cite{guanNoreferenceBlurAssessment2015}.

%The here implemented focus selection method integrates statistical concept into the Edge Model based Blur Metric (EMBM) proposed in \cite{guanNoreferenceBlurAssessment2015}. 

Following the Nucleus detection step (which is performed at the central focus level, $z=0$) we cut out a square region for each detected nucleus at all acquired focus levels. Each such cutout image is filtered with a small median filter of size $m\!\times\!m$ on each color channel to reduce noise.
%
%Let $\mathcal{P}$ denote 
This gives us a set of images $P_i$, $i = 1,\dots,N$, of an individual nucleus at the N consecutive z-levels.
%Having a set of smoothed patch $P_i$, $i = 1,\dots,N$, of an individual nucleus at N consecutive z-levels, 
We compute the difference of neighboring focus levels, $P'_{i} = P_{i+1} - P_{i}$, $i=1,\ldots,N\!-\!1$.
%\begin{equation}
%\label{eq:diffI}
%    \mathcal{P}' = \left\{ P'_{i} \mid P'_{i} = P_{i+1} - P_{i},\ i = 1,\dots, (N-1) %\right\}.
%\end{equation}
%
The variance, $\sigma^2_i$, is computed for each difference image $P_i'$:
%to form a SD sequence $\mathcal{SD}$:
\begin{equation*}
%\label{eq:SD}
    \textstyle %\mathcal{SD} = \Big\{\sigma_i \mid 
    \sigma^2_i = \frac{1}{M} \sum_{j=1}^{M}\left(p'_{ij}-\mu_i \right)^{2} \,,% \Big\},
\text{ where } \mu_i = \frac{1}{M} \sum_{j=1}^{M} p'_{ij} \,,
\end{equation*}
$M$ is the number of pixels in $P'_i$, and $p'_{ij}$ is the value of pixel $j$ in $P'_i$. 
Finally the level $l$ corresponding to the largest $\sigma^2_i$ is selected,
\begin{equation*}
    l=\argmax_{i=1,\ldots,N-1} \sigma^2_i\,.
\end{equation*}
To determine which of the two images in the pair $P'_l$ is in best focus, we use the EMBM method\cite{guanNoreferenceBlurAssessment2015} as a post selection step to choose which of images $P_l$ and $P_{l+1}$ to use.

\subsection{Classification}
\label{sec:method_C}

The final module of the pipeline is classification of the generated nucleus patches into two classes -- cancer and healthy.
%, based on their difference in texture \cite{jabaleeIdentificationMalignancyAssociatedChanges2018}. 
%
%Since ResNet was shown to be a preferable classifier architecture for a similar task,
Following recommendation from \cite{forslidDeepConvolutionalNeural2017}, we evaluate ResNet50 \cite{heDeepResidualLearning2015} as a classifier. We also include the more recent DenseNet201 \cite{huangDenselyConnectedConvolutional2018} architecture. In addition to random (Glorot-uniform) weight initialization, we also evaluate the two architectures using weights pre-trained on ImageNet.

% %We observe that the colors can sometimes distract the classifier from focusing on the texture, 
% The color information in this type of images is known to be less reliable than nucleus texture;
% we observe improved performance when relying on grayscale data compared to when using the full RGB information. 
% %we observe worse performance when using the RGB color information than when converting the images to grayscale.
% %To avoid distraction of the classifier by the color information, and we t
% Therefore we convert the images 
% %from RGB 
% to grayscale, $L = 0.299 R + 0.587 G + 0.114 B$. 

Considering that texture information is a key feature for classification\cite{jabaleeIdentificationMalignancyAssociatedChanges2018,wetzerTextureMatters2020}, the data is augmented without interpolation.
%to avoid any influence 
%on the relation 
%of neighboring pixels. 
During training, each sample is reflected with 50\% probability and rotated by a random integer multiple of \ang{90}.
%angle $\phi \in \{\ang{0},\ang{90},\ang{180},\ang{270}\}$.

%Each training sample is vertically or horizontally mirrored, and rotated by the multiple of \ang{90}. 
%resulting 8 times in number, including the original one. 
%Only one of the eight obtained versions of the sample is randomly chosen during training.

\section{Experimental Setup}
\label{sec:ExperimentalSetup}
%TensorFlow 1.14 is used in our implementation.

\subsection{Nucleus Detection}
\label{sec:ExperimentalSetup_ND}

The WSIs at the middle z-level ($z=0$) are used for nucleus detection.
Each WSI is split into an array of %JPEG 
$6496\!\times\!3360\!\times\!3$ sub-images using the Open Source tool \texttt{ndpisplit} \cite{deroulersAnalyzingHugePathology2013}.
%, with each sub-image of size (6496, 3360, 3). %The sub-images at the margin of the whole slide images are discarded due to different shapes and lack of information. 
The model is trained on 12 and tested on 2 sub-images (1014 resp.\ 119 nuclei) from Dataset~3.
The manually marked ground truth is dilated by a disk of radius $r\!=\!15$. 
%The images are originally of size (6496, 3360). But 
%Due to limitation of memory, 
All images, including ground truth masks, are resized to $1024\!\times\!512$ pixels, using area weighted interpolation. A Gaussian filter, $\sigma\!=\!1$, is applied to each ground truth mask providing the fuzzy ground truth $D$.

Each image is normalized by subtracting the mean and dividing by the standard deviation of the training set. 
%Some simple methods are also used for data augmentation: 
Images are augmented by random rotation in the range $\pm \ang{30}$, random horizontal and vertical shift within $30\%$ of the total scale, random zoom within the range of $30\%$ of the total size, and random horizontal and vertical flips. Nucleus detection does not need the texture details, so interpolation does not harm.
%
%To make the model easier to train, 
To improve stability of training,
batch normalization \cite{ioffeBatchNormalizationAccelerating2015} is added before each activation layer. Training is performed using RMSprop with mean squared error as loss function, learning rate $\alpha = 0.001$ and decay rate $\rho = 0.9$. %Limited by available memory, 
The model is trained with mini-batch size 1 for 100 epochs, the checkpoint with minimum training loss is used for testing.

\begin{figure}[tbp]
    \centering
    \includegraphics[width=0.45\textwidth]{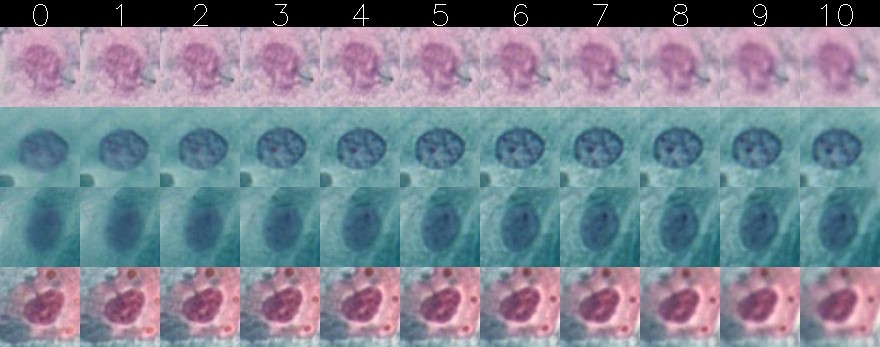}
    \caption{Example of focus sequences for experts to annotate}
    \label{fig:zeval}
\end{figure}

Performance of nucleus detection is evaluated on Dataset~3. %(Sect.~\ref{sec:ExperimentalSetup_data}) 
%and quantified in terms of Precision, Recall, and F1-score.
%Precision is the number of correctly detected nuclei divided by total number of detected nuclei. Recall is the number of correctly detected nuclei divided by total number of true nuclei. F1-score is the harmonic mean of precision and recall.
% , as defined in \eqref{eq:precision}, \eqref{eq:recall}, and \eqref{eq:f1}. %
%
% \begin{equation}
% \label{eq:precision}
% \begin{split}
%     Precision 
%     &= \frac{Number\ of\ correctly\ detected\ nuclei}{Total\ number\ of\ detected\ nuclei} \\
%     &= \frac{TP}{TP + FP}    
% \end{split}
% \end{equation}
%
% \begin{equation}
% \label{eq:recall}
% \begin{split}
%     Recall
%     &= \frac{Number\ of\ correctly\ detected\ nuclei}{Total\ number\ of\ true\ nuclei} \\
%     &= \frac{TP}{TP + FN}    
% \end{split}
% \end{equation}
%
% \begin{equation}
% \label{eq:f1}
%     F1\text{-score} = 2 \cdot \frac{Precision \cdot Recall}{Precision + Recall}
% \end{equation}
%
%
%The number of correctly detected nuclei (TP) is defined as follows:
%according to Algorithm \ref{alg:TPcount}: 
A detection is considered correct if its closest ground truth nucleus is within the cropped patch
%closer to a ground truth nucleus than a Chebyshev distance $d=40$\,pixels 
\textit{and} that ground truth nucleus has no closer detections
%it is the closest detection of that ground truth nucleus 
(s.t. one true nucleus is paired with at most one detection).
%For each detected nucleus, 
%the closest ground truth nucleus, and the closest detection to that ground truth are found. If they match and the Chebyshev distance between current detection and its closest ground truth is less than or equal to the detection window size $w$, the current detection is a true positive. 
%In this experiment, we use $d = 80$.

%\begin{algorithm}[ht]
%\SetAlgoLined
%\LinesNumbered
%\KwIn{$S$ - set of detected nuclei positions, $T$ - set of true nuclei %positions, $w$ - size of detection window}
%\KwOut{$n_\text{{TP}}$ - number of correctly detected nuclei}
%$n_\text{{TP}} \leftarrow 0$\;
%\For{$s \in S$}{
    %$ t \leftarrow \mathop{\argmin}_{x \in T} d(x, s) $\;
    %$ s' \leftarrow \mathop{\argmin}_{y \in S} d(y, t) $\;
    %\If{$ s = s'\ \land\ D_{Chebyshev}(s, t) \leq w/2 $}{
        %$n_\text{{TP}} \leftarrow n_\text{{TP}} + 1$\;
    %}
%}
    %\caption{True positive counting}
    %\label{alg:TPcount}
%\end{algorithm}

\subsection{Focus Selection}
\label{sec:ExperimentalSetup_FS}

100 detected nuclei are randomly chosen from the two test sub-images (Dataset~3). Every nucleus is cut to an $80\!\times\!80\!\times\!3$ patch for each of the 11 z-levels. For EMBM method the contrast threshold of a salient edge is set to $c_T=8$, following\cite{guanNoreferenceBlurAssessment2015}.

To evaluate the focus selection, 8 experts are asked to choose the best of the 11 focus-levels for each of the 100 nuclei (Fig.~\ref{fig:zeval}). %Assuming the focus quality is changing monotonically as the z-offset changes, two different ground truths are defined: 
%range ground truth and median ground truth $GT_{median}$. 
%The {\it range ground truth} is the interval between minimum label $GT_{min}$ and maximum label $GT_{max}$. 
The median of the 8 assigned labels is used as true best focus, $l_{GT}$. 
A predicted focus level $l$ is considered accurate enough
%a true positive w.r.t. range ground truth if 
%$p_f \in [GT_{min},\ GT_{max}]$,
% $$\min(1, GT_{min} - \delta_range),\ \max(GT_{max} + \delta_range, N)$$   % new range accuracy
%and is a true positive w.r.t. $GT_{median}$ 
if $l \in [l_{GT}-2,l_{GT}+2]$. 
%where $\delta$ is the allowed deviation (set to 2 in this experiment).
%The performance of focus selection is evaluated on Dataset 3. %(Sect.~\ref{sec:ExperimentalSetup_data}).
%and quantified in terms of median accuracy and range accuracy.

\subsection{Classification}
\label{sec:ExperimentalSetupC}

The classification model is evaluated on Dataset 1 as a benchmark, and then on Dataset 2, to evaluate effectiveness of the nucleus detection and focus selection modules in comparison with the performance on Dataset 1. The model is also run on Dataset 3 to validate generality of the pipeline. Datasets are split on a patient level; no cell from the same patient exists in both training and test sets. On Dataset 1 and 2, three-fold validation is used, following
%Table 1
\cite{forslidDeepConvolutionalNeural2017}. On Dataset 3, two-fold validation is used. 
Our trained nucleus detector with threshold $T = 0.59$ (best-performing in Sec.~\ref{sec:ResultsDiscussion ND}) is used for Dataset 2 and 3 to generate nucleus patches.
Some cells in Dataset 2 and 3 lie outside the $\pm \SI{2}{\um}$ imaged z-levels, and the best focus is still rather blurred. We use the EMBM to exclude the most blurred ones. Cell patches with an EMBM score $<0.03$ are removed, leaving $68509$ cells for Dataset 2 and $130521$ for Dataset 3.

% Using a random forest classifier \cite{jabaleeIdentificationMalignancyAssociatedChanges2018} reliably detected Malignancy associated changes (MAC) in histologically normal (normal-appearing) oropharyngeal epithelial cells located in tissue samples adjacent to a tumor. This encourages usage of MAC as a noninvasive means of detecting early-stage oropharyngeal tumors. 
% %Probably no space for further discussion:
% In addition, feature selection procedure used in~\cite{jabaleeIdentificationMalignancyAssociatedChanges2018} confirms earlier observations that texture features are most significant for reliable MAC and cancer detection.

\begin{figure}[tbp]
    \centering
    \subfloat[\!Performance for different thresholds $T$\!]{
        \includegraphics[width=0.4\textwidth]{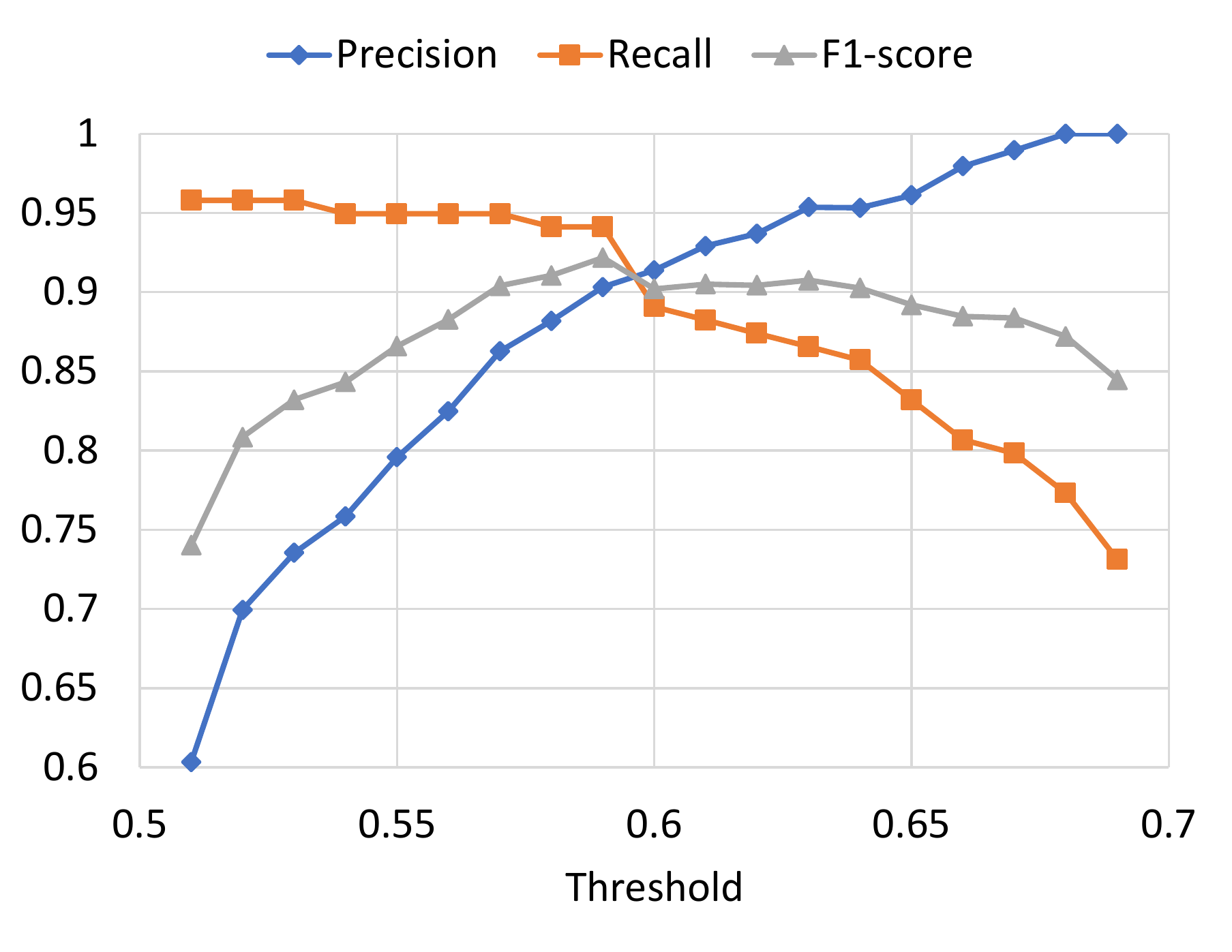}
        \label{fig:DetectionPerformance}}\hfil
    \subfloat[Precision-recall curve]{
        \includegraphics[width=0.4\textwidth]{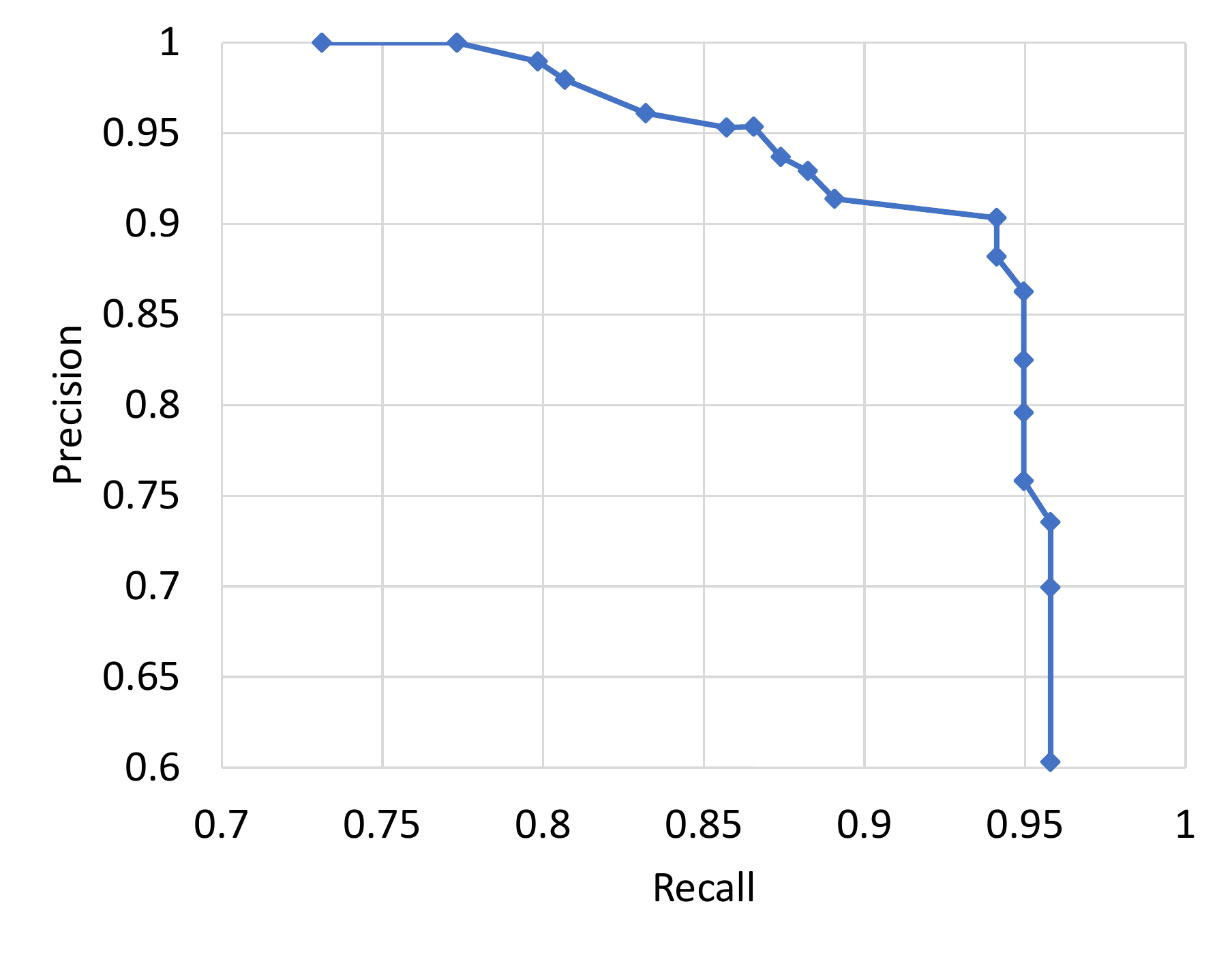}
        \label{fig:PRcurve}}

    \caption{Results of nucleus detection}
    \label{fig:detectionresult}
\end{figure}

We use Adam optimizer, cross-entropy loss and parameters as suggested in \cite{kingmaAdamMethodStochastic2014}, i.e., initial learning rate $\alpha=0.001$, $\beta_1=0.9$, $\beta_2=0.999$. 
$10\%$ of the training set is randomly chosen as validation set. 

% For ResNet50, we observe improved performance when relying on grayscale data compared to when using the full RGB information. Therefore we convert the input images to grayscale, $L = 0.299 R + 0.587 G + 0.114 B$. 
When using models pre-trained on ImageNet, since the weights require three input channels, the grayscale images from Dataset 1 are duplicated into each channel. Pre-trained models are trained (fine-tuned) for 5 epochs. The learning rate is scaled by 0.4 every time the validation loss does not decrease compared to the previous epoch. The checkpoint with minimum validation loss is saved for testing. 

A slightly different training strategy is used when training from scratch. 
%The grayscale images from Dataset 1 are pre-processed in the same way as for the pre-trained models. 
ResNet50 models are trained with mini-batch size 512 for 50 epochs on Dataset 2 and 3, and with mini-batch size 128 for 30 epochs on Dataset 1, since it contains fewer samples. Because DenseNet201 takes larger GPU memory, mini-batch sizes are set to 256 on Dataset 2 and 3. To mitigate overfitting, DenseNet201 models are trained for only 30 epochs on Dataset 2 and 3, and 20 epochs on Dataset 1. 
When the validation loss has not decreased for 5 epochs, the learning rate is scaled by 0.1. Training is stopped after 15 epochs of no improvement. The checkpoint with minimum validation loss is saved for testing.

%The classification performance is quantitatively evaluated by several metrics: accuracy, class-weighted precision, recall and F1-score. Cohen's kappa coefficient is also evaluated as a function of defocus level on fold 1 of Dataset 2. Furthermore, to illustrate the practicability of this pipeline, percentages of cells classified as tumor cells are plotted for each glass on all three datasets.

\section{Results and Discussion}
\label{sec:ResultsDiscussion}

\subsection{Nucleus Detection}
\label{sec:ResultsDiscussion ND}

Results of nucleus detection are presented in Fig.~\ref{fig:detectionresult}. Fig.~\ref{fig:DetectionPerformance} shows Precision, Recall, and F1-score as the detection threshold $T$ varies in $[0.51,0.69]$. At $T = 0.59$, F1-score reaches 0.92, with Precision and Recall being 0.90 and 0.94 respectively.
%Precision reaches 1.00 for $T = 0.68$, with Recall being 0.77, as shown in Fig.~\ref{fig:PRcurve}.
Using $T=0.59$, 94,685 free lying nuclei are detected in Dataset 2 and 138,196 in Dataset 3.

%The processing time tested on NVIDIA GeForce GTX 1060 Max-Q is also recorded during the prediction step. 
The inference takes $\SI{0.17}{s}$ to generate a density map $D'$ of size $1024\!\times\!512$ on an NVIDIA GeForce GTX 1060 Max-Q. To generate a 
density map of the same size based on the sliding window approach (Table 4 of \cite{hofenerDeepLearningNuclei2018}), takes $\SI{504}{s}$.
 
%at the best speed $6.84 MP/s$ of sliding window approach reported in Table 4 of \cite{hofenerDeepLearningNuclei2018}, generating a density map $D'$ of the same size takes $504.49 s$.

% the sliding window approach takes $298 s$, (similar to the speed reported in \cite{hofenerDeepLearningNuclei2018}) at step-size of 10 pixels, which even generates a 2.6 times smaller density map D'(x) of size (634, 320).

\subsection{Focus Selection}
\label{sec:ResultsDiscussion_FS}

\begin{figure}[tb]
    \centering
    \includegraphics[width=0.45\textwidth]{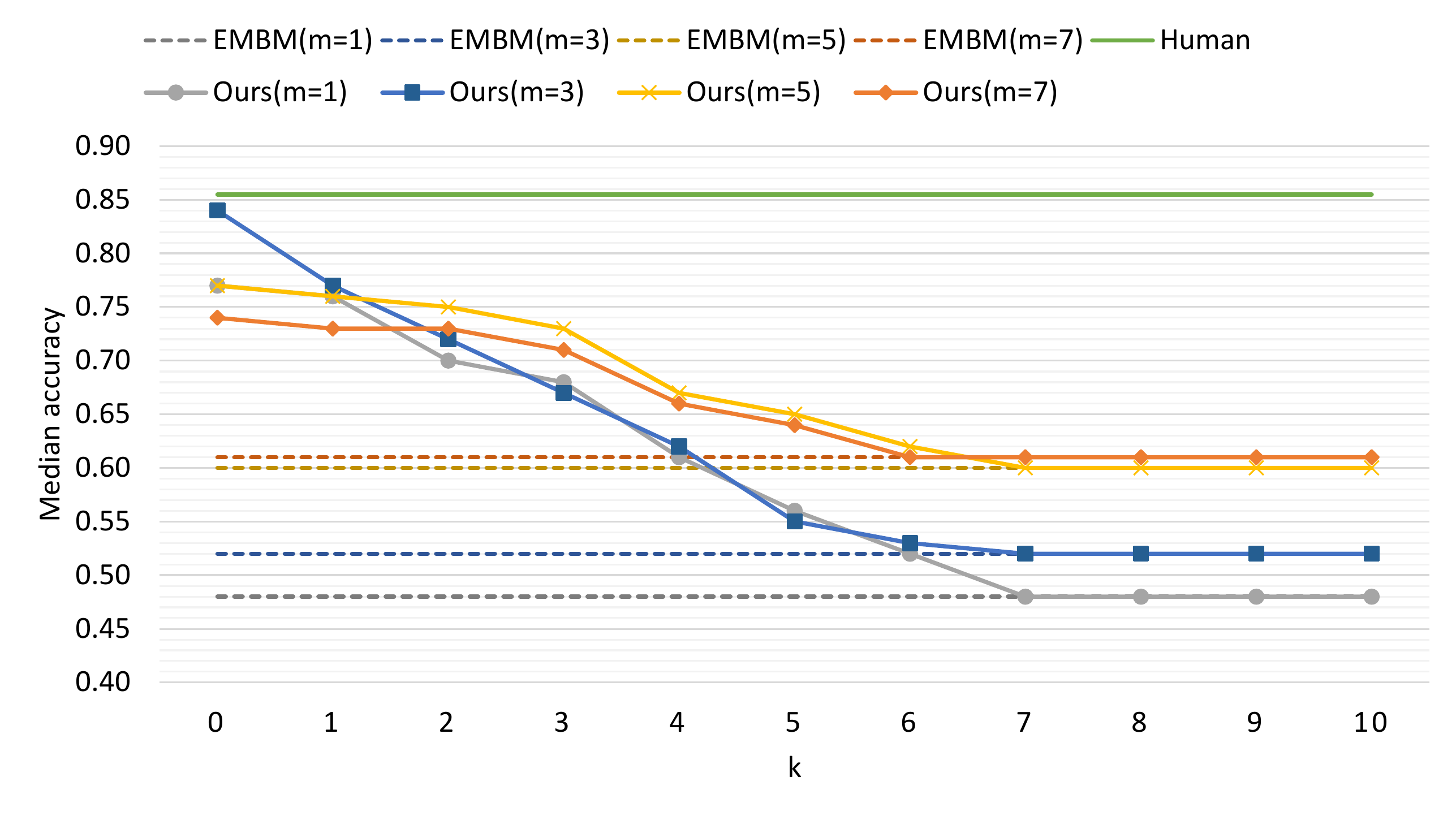}
    \caption{Accuracy of focus selection}
    \label{fig:k}
\end{figure}

Performance of the focus selection is presented in Fig.~\ref{fig:k}. The ``human'' performance is the average of the experts, using a leave-one-out approach.
%Each expert's performance is computed based on the statistics of the rest.
We plot performance when using EMBM to select among the $2(k+1)$ levels closest to our selected pair $l$; for increasing $k$ the method approaches EMBM.

It can be seen that EMBM alone
%without pre-processing 
does not achieve satisfying performance on this task. 
%This may be due to the difficulty of focus selection in microscopy images, or because out-of-focus images contain not only blur but also artifacts and background noise. 
Applying a median filter improves the performance somewhat.
% by avoiding measuring edges of high-frequency pixel-wise noise. 
%Artifacts show less variation w.r.t. focus depth than nuclear texture and
Our proposed method performs very well on the data and is essentially at the level of a human expert (accuracy 84\% vs. 85.5\%, respectively) using $k=0$ and a $3\!\times\!3$ median filter.

% \begin{table}[tbp]
%     \centering
%     \caption{Focus selection performance}
%     \label{tab:res_focus}
%     \resizebox{0.48\textwidth}{!}{%
%     \begin{tabular}{|c|c|c|c|}
%         \hline
%         \multicolumn{2}{|c|}{\textbf{Pre-processing}} & \multicolumn{2}{c|}{\textbf{Performance}} \\ \hline
%         Median filter size   & SD limitation         & Median accuracy      & Range accuracy     \\ \hline
%         None                 & FALSE                 & 0.48                 & 0.58               \\ \hline
%         (3, 3)               & FALSE                 & 0.52                 & 0.63               \\ \hline
%         (5, 5)               & FALSE                 & 0.60                 & 0.69               \\ \hline
%         (7, 7)               & FALSE                 & 0.61                 & 0.67               \\ \hline
%         None                 & TRUE                  & 0.76                 & 0.83               \\ \hline
%         (3, 3)               & TRUE                  & \textbf{0.77}        & 0.82               \\ \hline
%         (5, 5)               & TRUE                  & 0.76                 & \textbf{0.89}      \\ \hline
%         (7, 7)               & TRUE                  & 0.73                 & 0.81               \\ \hline
%         \multicolumn{2}{|c|}{Human performance}      & 0.855                & 0.865              \\ \hline
%     \end{tabular}%
%     }
% \end{table}

\subsection{Classification}
\label{sec:ResultsDiscussion_C}
 
\begin{figure*}[tb]
    \captionsetup[subfigure]{justification=raggedright}
    \centering
    \subfloat[Dataset 1, manual cell selection, 10274 cells, ResNet50]{\includegraphics[width=0.33\textwidth]{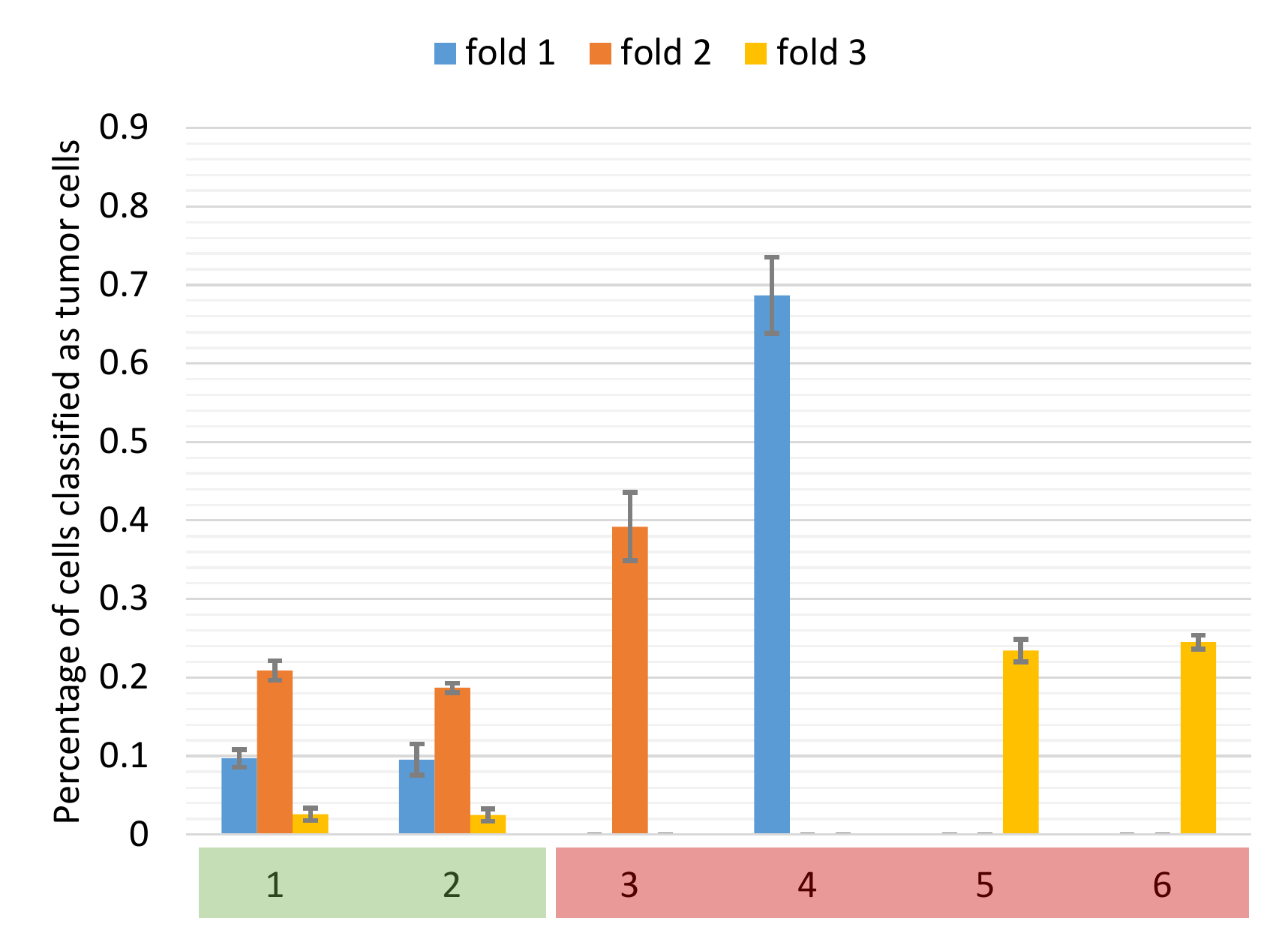}\label{fig:classification11}}\hfil
    \subfloat[Dataset 2, fully automatic pipeline, 68509 cells, ResNet50]{\includegraphics[width=0.33\textwidth]{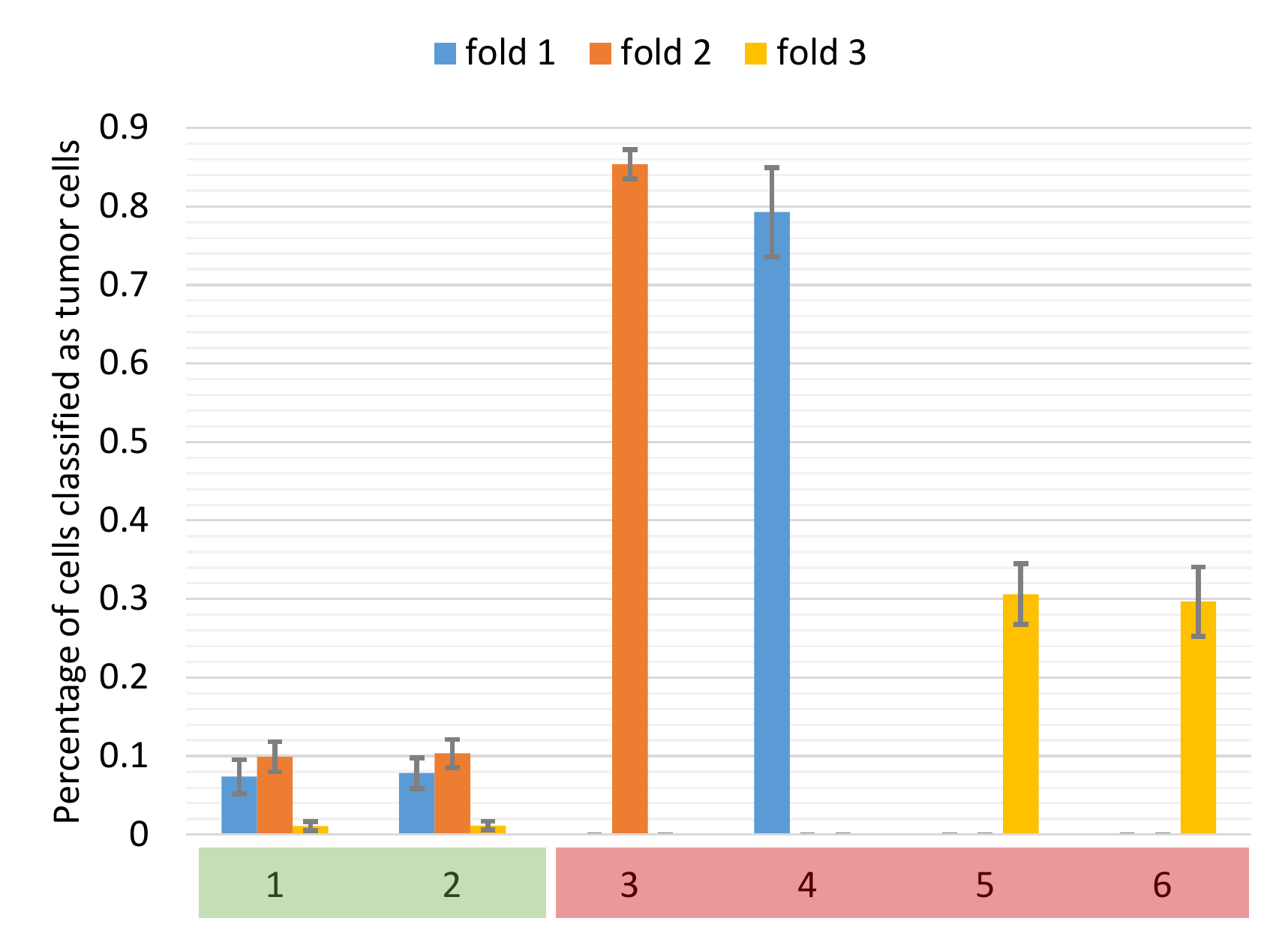}\label{fig:classification12}}\hfil
    \subfloat[Dataset 3 (LBC), fully automatic, 130521 cells, ResNet50]{\includegraphics[width=0.33\textwidth]{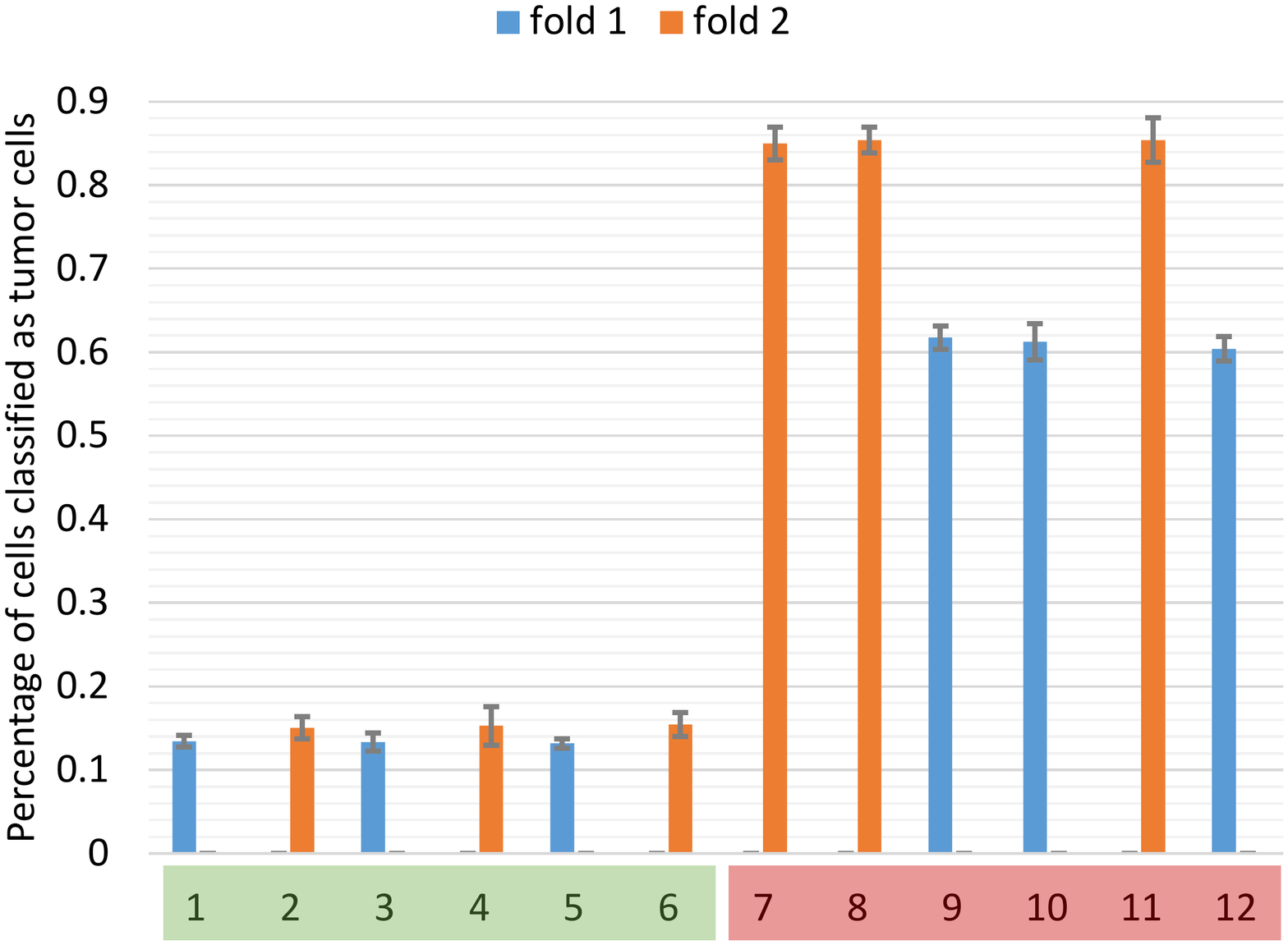}\label{fig:classification13}}
    
% \captionsetup[subfigure]{justification=raggedright} %To not make huge space between "Dataset" and number
%     \subfloat[Dataset 1, ResNet50(pre-trained)]{\includegraphics[width=0.33\textwidth]{images/plot_ResNet50_pre1_dataset_1.pdf}\label{fig:classification21}}\hfil
%     \subfloat[Dataset 2, ResNet50(pre-trained)]{\includegraphics[width=0.33\textwidth]{images/plot_ResNet50_pre1_dataset_2.pdf}\label{fig:classification22}}\hfil
%     \subfloat[Dataset 3, ResNet50(pre-trained)]{\includegraphics[width=0.33\textwidth]{images/plot_ResNet50_pre1_dataset_3.pdf}\label{fig:classification23}}

% \captionsetup[subfigure]{justification=raggedright} %To not make huge space between "Dataset" and number
%     \subfloat[Dataset 1, DenseNet201]{\includegraphics[width=0.33\textwidth]{images/plot_DenseNet201_pre0_dataset_1.pdf}\label{fig:classification31}}\hfil
%     \subfloat[Dataset 2, DenseNet201]{\includegraphics[width=0.33\textwidth]{images/plot_DenseNet201_pre0_dataset_2.pdf}\label{fig:classification32}}\hfil
%     \subfloat[Dataset 3, DenseNet201]{\includegraphics[width=0.33\textwidth]{images/plot_DenseNet201_pre0_dataset_3.pdf}\label{fig:classification33}}

\captionsetup[subfigure]{justification=raggedright} %To not make huge space between "Dataset" and number
    \subfloat[Dataset 1, DenseNet201(pre-trained)]{\includegraphics[width=0.33\textwidth]{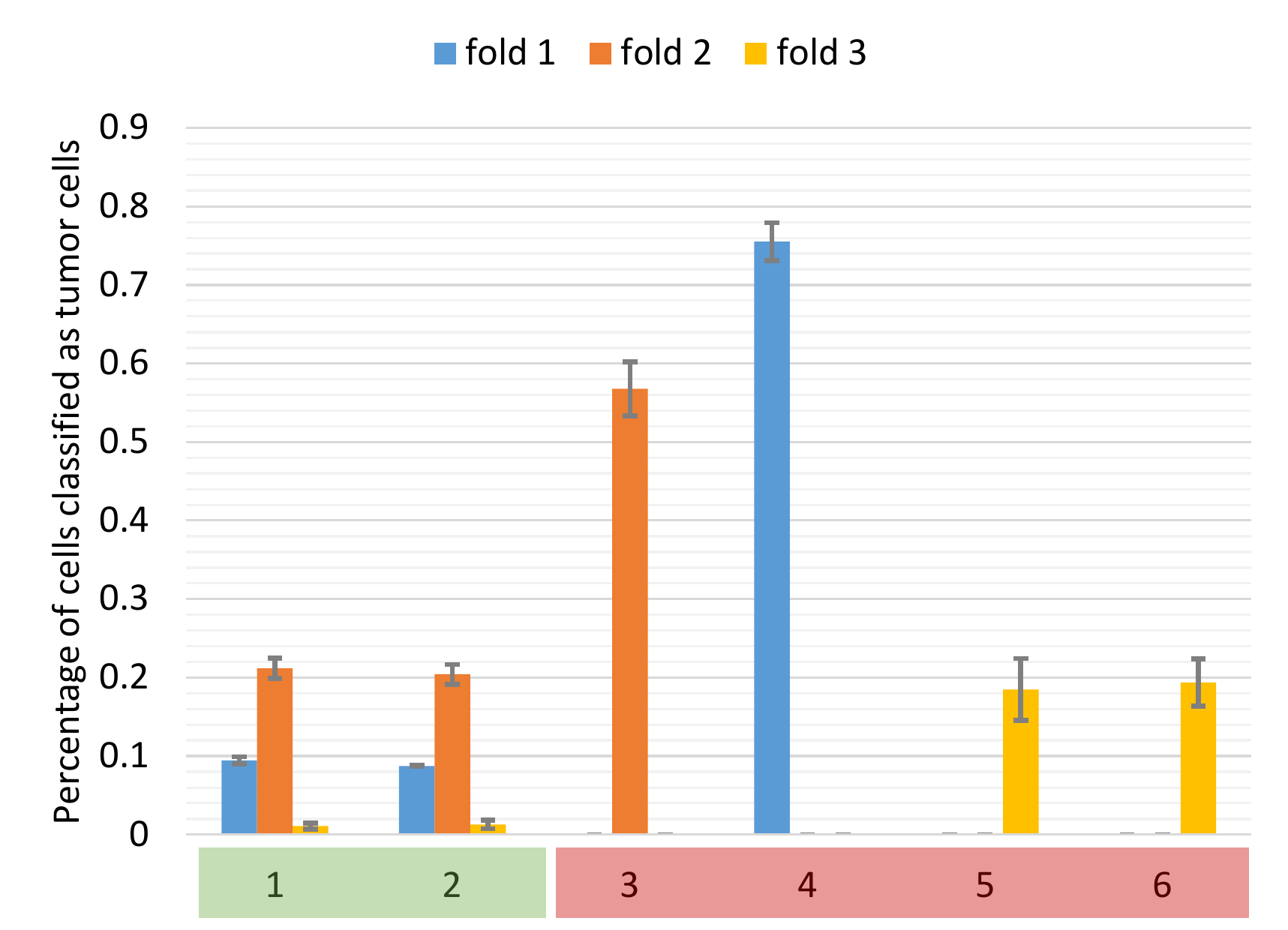}\label{fig:classification41}}\hfil
    \subfloat[Dataset 2, DenseNet201(pre-trained)]{\includegraphics[width=0.33\textwidth]{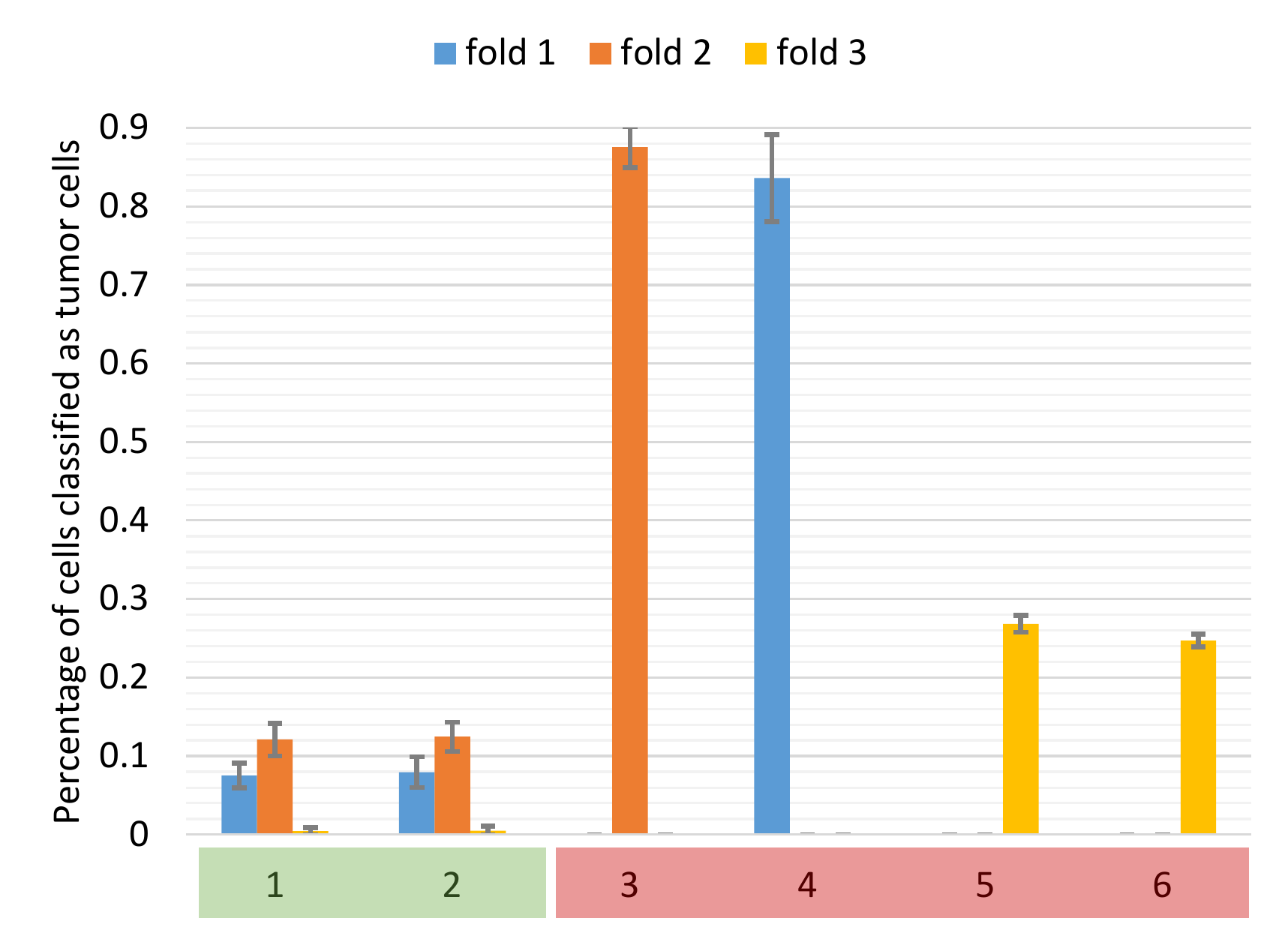}\label{fig:classification42}}\hfil
    \subfloat[Dataset 3, DenseNet201(pre-trained)]{\includegraphics[width=0.33\textwidth]{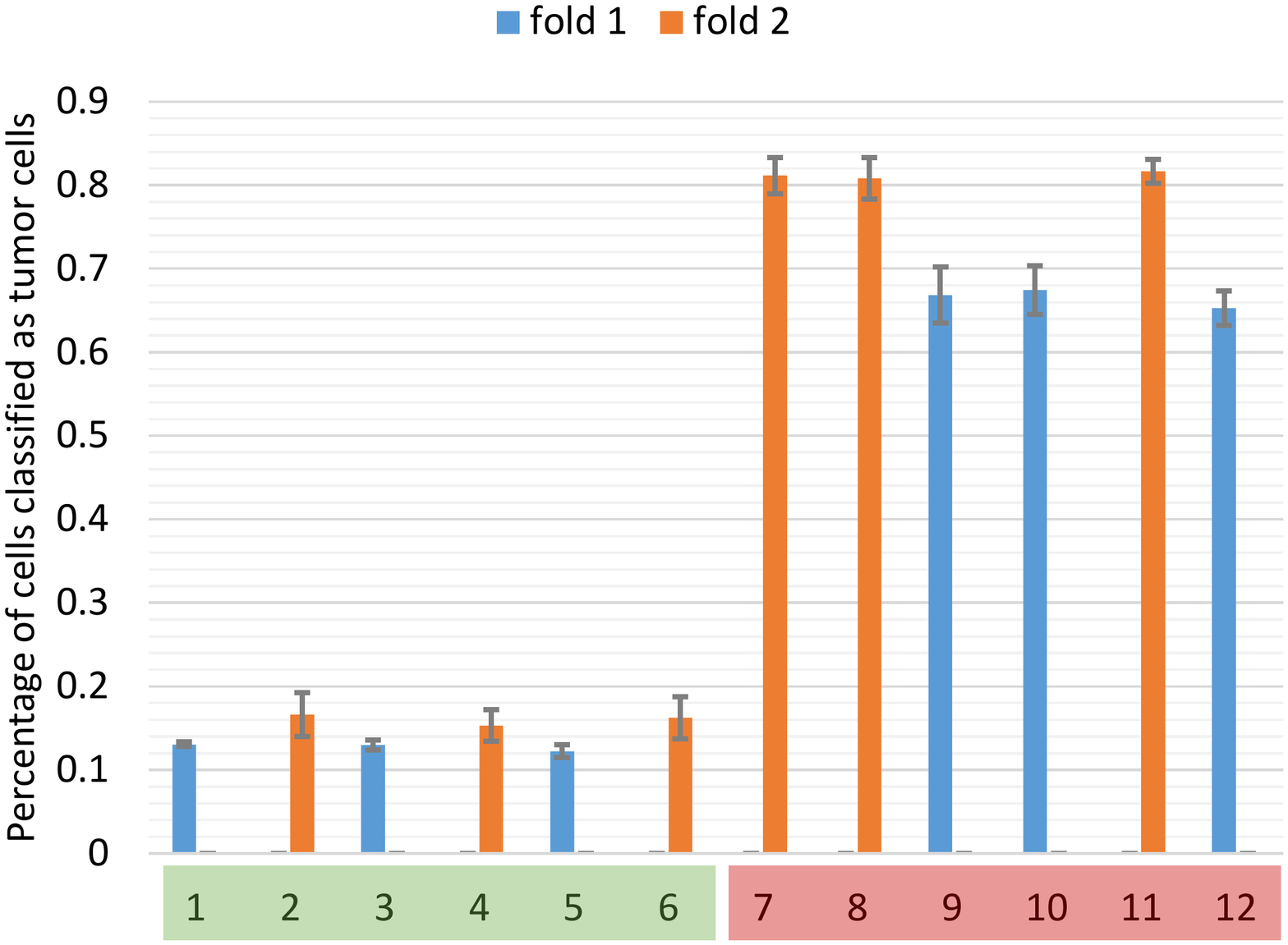}\label{fig:classification43}}
    \caption{Cell classification results per microscope slide; green samples (bars to the left) are healthy, red samples (bars to the right) are from cancer patients. ResNet50 is used for (a)-(c) and DenseNet201 pre-trained on ImageNet is used for (d)-(f).} %, g-i are trained with DenseNet201 pre-trained on ImageNet.
    \label{fig:classification}
\end{figure*}

Classification performance is presented in Table~\ref{tab:classificationresults} and Fig.~\ref{fig:classification}. The two architectures (ResNet50 and DenseNet201) perform more or less equally well. Pre-training seems to help a bit for the smaller Dataset 1, whereas for the larger Datasets 2 and 3 no essential difference is observed.
Results on Dataset 2 are consistently better than on Dataset 1. 
This confirms effectiveness of the nucleus detection and focus selection modules; by using more nuclei (from the same samples) than those manually selected, improved performance is achieved. The results on Dataset 3 indicate that the pipeline generalizes well to liquid-based images. We also observe that our proposed pipeline is robust w.r.t. network architectures and training strategies of the classification.

% Comparing pre-trained DenseNet201 with ResNet50, although better results are achieved on Dataset 3 and percentages of cells classified as tumor cells are higher in some slides of Dataset 2 (Fig.~\ref{fig:classification12}-\ref{fig:classification13}, \ref{fig:classification42}-\ref{fig:classification43}), the two classes of slides become inseparable on Dataset 1 (Fig.~\ref{fig:classification11} and \ref{fig:classification41}). Dataset 1 contains much fewer samples and a more advanced network may overfit to the training set. 
% %These results indicate that our proposed pipeline is robust to network architectures and can avoid human bias in selecting cell patches. 
% The pipeline generalizes better on large data when using more advanced network.
% % For smaller training sets (Dataset 1) we observe somewhat more stable performance using pre-training on ImageNet (Tab.~\ref{tab:classificationresults}), whereas for larger datasets we do not see any such improvement.
% We observe better generalization and more stable performance using pre-training on ImageNet (Tab.~\ref{tab:classificationresults}).

\begin{figure}[tbp]
    \centering
    \includegraphics[width=0.45\textwidth,trim={0 2mm 0 2mm},clip]{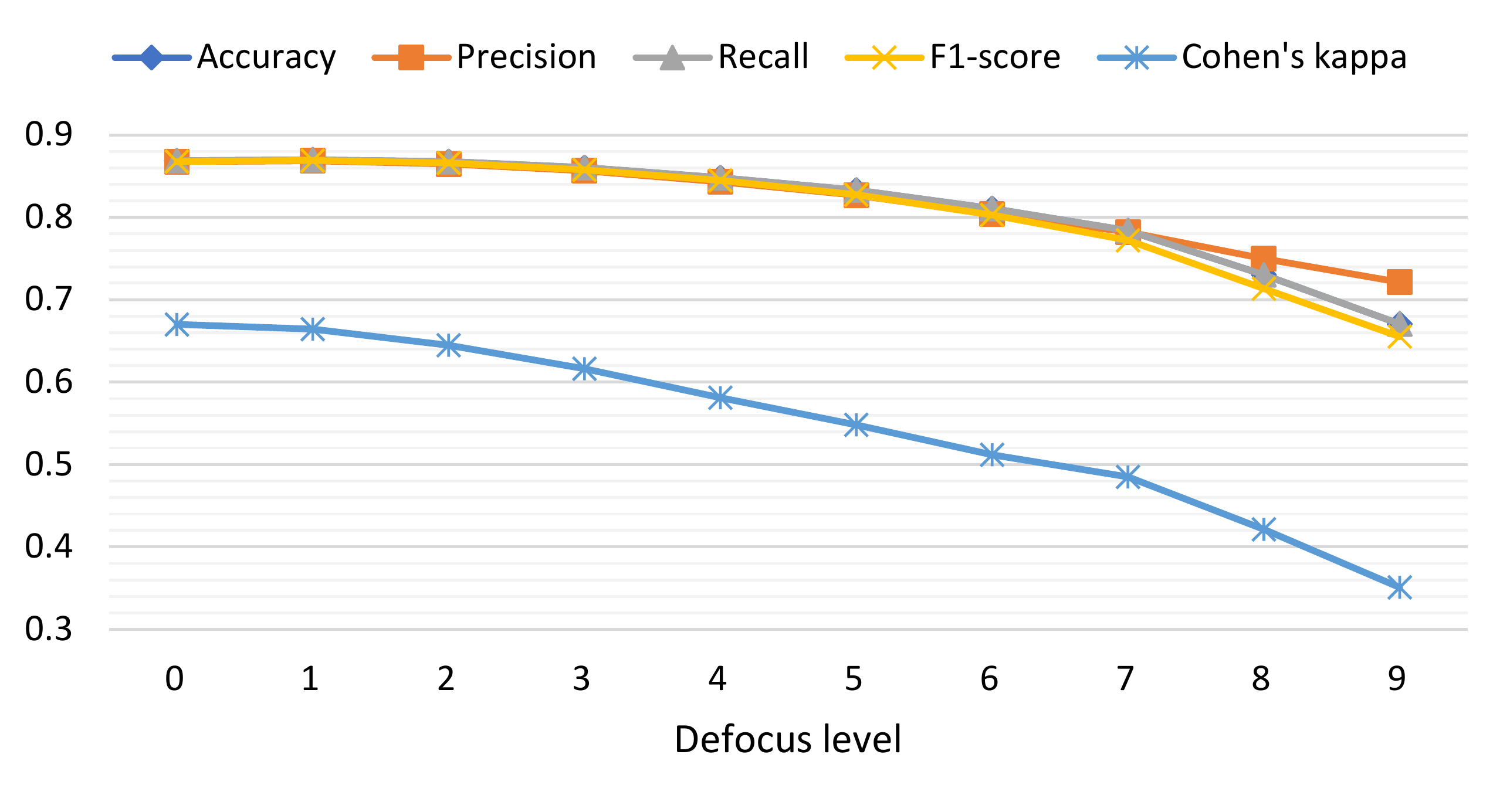}
    \caption{The impact of defocused testset on Dataset 2, fold 1 (ResNet50)}
    \label{fig:defocus}
\end{figure}

In Fig.~\ref{fig:defocus} we plot how classification performance decreases when nuclei are intentionally selected $n$ focus levels away from the detected best focus. The drop in performance as we move away from the detected focus confirms the usefulness of the focus selection step.
%Cohen's kappa score decreases fast as the defocus level of test-set increases. This indicates the necessity of focus selection step in the screening pipeline. 
%The bounce at the defocus level 10 could be explained by the bias due to the statistical lack of samples. There are only $1019$ samples at level 10, compared to $24031$ at level 1.

If aggregating the cell classifications over whole microscopy slides, as show in Fig.~\ref{fig:classification}, comparing Fig.~\ref{fig:classification11}-\ref{fig:classification12} and Fig.~\ref{fig:classification41}-\ref{fig:classification42}, we observe that the non-separable slides 1, 2, 5, and 6 in Dataset 1 become separable in Dataset 2. Global thresholds can be found which accurately separate the two classes of patients in both datasets processed by our pipeline.

% Comparing
% the peformance on Dataset 1 and 2, 
% %Fig.~\ref{fig:classification12} with Fig.~\ref{fig:classification11}, in each fold, the values of the glasses from the same class are closer, and there are larger gaps between classes. The values in Dataset 2 also show much smaller standard deviation in Table \ref{tab:classificationresults}. 
% we conclude that the presented pipeline can reduce human bias, and make the classification easier and more stable.

% \usepackage{multirow}
\begin{table}[tbp]
    \centering
    \caption{Classification performance. The best F1-score for each dataset is presented in bold.}
    \label{tab:classificationresults}
    \resizebox{0.48\textwidth}{!}{%
    \begin{tabular}{ccccccc}
        \hline

        \textbf{Dataset} & \textbf{Network} & \textbf{Accuracy} & \textbf{Precision} & \textbf{Recall} & \textbf{F1-score} \\ \hline
        \multirow{4}{*}{1} & ResNet50 & 70.5$\pm$0.5 & 63.1$\pm$1.2 & 34.8$\pm$1.4 & 44.8$\pm$1.3 \\
         & ResNet50(pre-trained) & 72.0$\pm$0.9 & 66.4$\pm$2.0 & 37.5$\pm$2.0 & \textbf{48.0$\pm$2.1} \\
         & DenseNet201 & 70.4$\pm$0.5 & 63.1$\pm$1.8 & 33.8$\pm$0.9 & 44.0$\pm$0.7 \\
         & DenseNet201(pre-trained) & 70.6$\pm$0.7 & 63.4$\pm$1.6 & 34.3$\pm$1.7 & 44.5$\pm$1.8 \\ \hline
        \multirow{4}{*}{2} & ResNet50 & 74.4$\pm$1.9 & 83.3$\pm$2.9 & 46.3$\pm$3.8 & 59.5$\pm$3.8 \\
         & ResNet50(pre-trained) & 74.0$\pm$0.1 & 83.9$\pm$0.5 & 44.6$\pm$0.7 & 58.2$\pm$0.5 \\
         & DenseNet201 & 75.4$\pm$0.8 & 84.3$\pm$1.5 & 48.3$\pm$1.1 & \textbf{61.4$\pm$1.3} \\
         & DenseNet201(pre-trained) & 73.3$\pm$0.7 & 81.7$\pm$2.8 & 44.4$\pm$0.3 & 57.5$\pm$0.6 \\ \hline
        \multirow{4}{*}{3} & ResNet50 & 81.6$\pm$0.7 & 71.7$\pm$1.2 & 73.8$\pm$0.9 & \textbf{72.8$\pm$1.0} \\
         & ResNet50(pre-trained) & 81.3$\pm$1.5 & 72.1$\pm$3.0 & 71.6$\pm$0.6 & 71.8$\pm$1.8 \\
         & DenseNet201 & 81.3$\pm$0.5 & 71.4$\pm$0.7 & 73.0$\pm$0.8 & 72.2$\pm$0.7 \\
         & DenseNet201(pre-trained) & 81.5$\pm$1.3 & 71.2$\pm$2.4 & 74.5$\pm$2.4 & 72.8$\pm$1.9 \\ \hline
        \end{tabular}%
        }
        % \vspace*{-5mm}
\end{table}

% The inference time is tested on Intel i9-7980XE CPU @ 2.60 GHz and a GPU cluster consisting one NVIDIA TITAN Xp, one TITAN V and one GeForce RTX 2080 Ti. For each whole slide image, the average computation time is about $17 min$ for nucleus detection, including inference and cell patch generation; the average focus selection time is about $77 min$ $7.8 s$, paralleled with 32 processes only on CPU; classification takes only about $1 s$.

\section{Conclusion}
\label{sec:Conclusion}

This work presents a complete fully automated pipeline for oral cancer screening on whole slide images; source code (utilizing TensorFlow 1.14) is shared as open source. The proposed focus selection method performs at the level of a human expert and significantly outperforms EMBM. The pipeline can provide fully automatic inference for WSIs within reasonable computation time. It performs well for smears as well as liquid-based slides.

%, and is shown to surpass the classification performance which uses human selected nuclei from the same slides.

Comparing the performance on Dataset 1, using human selected nuclei and Dataset 2, using computer selected nuclei from the same microscopy slides,
%Fig.~\ref{fig:classification12} with Fig.~\ref{fig:classification11}, in each fold, the values of the glasses from the same class are closer, and there are larger gaps between classes. The values in Dataset 2 also show much smaller standard deviation in Table \ref{tab:classificationresults}. 
we conclude that the presented pipeline can reduce human workload while at the same time make the classification easier and more reliable.

%The focus selection module still has the potential to accelerate with the DeepFocus method proposed in \cite{senarasDeepFocusDetectionOutoffocus2018}. And some hyper-parameters can be tuned to improve the performance on the liquid-based slides.

%\section*{Acknowledgments}
%We thank Tor Halle, Uppsala University Hospital, for help with slide scanning.

\bibliographystyle{IEEEbib}
\bibliography{ref}

% For citations of references, we prefer the use of square brackets
% and consecutive numbers. Citations using labels or the author/year
% convention are also acceptable. The following bibliography provides
% a sample reference list with entries for journal
% articles~\cite{ref_article1}, an LNCS chapter~\cite{ref_lncs1}, a
% book~\cite{ref_book1}, proceedings without editors~\cite{ref_proc1},
% and a homepage~\cite{ref_url1}. Multiple citations are grouped
% \cite{ref_article1,ref_lncs1,ref_book1},
% \cite{ref_article1,ref_book1,ref_proc1,ref_url1}.
% %
% % ---- Bibliography ----
% %
% % BibTeX users should specify bibliography style 'splncs04'.
% % References will then be sorted and formatted in the correct style.
% %
% % \bibliographystyle{splncs04}
% % \bibliography{mybibliography}




% LNCS Homepage, \url{http://www.springer.com/lncs}. Last accessed 4
% \end{thebibliography}
\end{document}